# FEASIBILITY OF THE DAFNE-LINAC ENERGY DOUBLING

Roberto Boni, INFN – Laboratori Nazionali di Frascati,
Via E. Fermi 40, 00044 Frascati (Italy)

*Abstract*

The energy doubling of the Frascati DAΦNE Φ-Factory can be accomplished either by ramping the energy of the rings from the initial 510 MeV of the injected beam or by upgrading the present Linac energy to > 1.02 GeV so to inject the beam at full energy. The choice between the two options is not the goal of this paper, which, instead, presents some technical solutions, which permit to re-use all the components of the DAFNE electron-positron Linac.

## THE FRASCATI LINAC

It consists mainly of two parts:
1) The low-energy/high-current section made up of five S-band SLAC-type 3 meters accelerating structures. The maximum achievable energy, on the Positron Converter (PC), is 250 MeV at zero electron current. In real operation, with 10 nsec, 5 amperes electron bunches at the PC, the energy conversion is 220 MeV.
2) The high-energy/low-current section made up of ten SLAC-type accelerating structures. It can accelerate the positron bunches emerging by the PC, up to the maximum energy of 550 MeV.

The SLAC-type structures are travelling wave (TW) constant gradient (CG) units and are fed by sledded RF pulses. Four RF stations provide the power by means of 45 MW klystrons. A sketch of the DAFNE Linac RF layout is shown in figure 1.

The main parameters of the DAFNE Linac are reported in table 1.

Table 1: DAFNE Linac Parameters

|  | DESIGN | OPERATIONAL |
|---|---|---|
| e- final energy | 800 MeV | 510 MeV |
| e+ final energy | 550 MeV | 510 MeV |
| Linac frequency | 2856 MHz | 2856 MHz |
| e+ conversion energy | 250 MeV | 220 MeV |
| beam pulse rep. rate | 1÷50 Hz | 1÷50 Hz |
| macropulse length | 10 nsec | 1÷10 nsec |
| Gun current | 8 A | 8 A |
| e- current on P.C. | 5 A | 5.2 A |
| max output e- current | > 150 mA | 350 mA |
| max output e+ current | 36 mA | 85 mA |

The klystron gallery footprint is shown in figure 2.

A drift space of about 15 meters is available at the linac output for the installation of other accelerating structures. Additional space is available to house even more structures, in the straight tube after the spectrometer magnet and in the transfer line tunnel before entering the main ring hall.

## DAFNE LINAC UPGRADING

The extra energy needed to inject at full energy in DAFNE2 is > 510 MeV.
Two solutions can be proposed to increase the energy
a) addition of new accelerating structures without any modification of the present machine;
b) addition of new accelerating structures with a new layout of the linac to increase the energy gain of the existing SLAC sections.

### The SLAC structures features

In TW/CG accelerating structures, the voltage $V_b$ (beam loading) induced by a bunch of peak current $I_0$ and duration $t_b$, very short with respect to the filling time $t_f$, is given by the following expression [1,2]:

$$V_b = \frac{R_{sh} I_0 L}{2(1-e^{-2\tau})}\left[1 - e^{-2\tau x_b} - 2\tau x_b e^{-2\tau}\right] \quad (1)$$

with: $R_{sh}$ = shunt impedance (MΩ/m)
   $L$ = structure length (m)
   $\tau$ = attenuation constant (Np)
 $x_b = t_b/t_f$ (defined in Figure 3).

In the SLAC structures, the above parameters are:

$\tau$ = 0.57 Neper
$R_{sh} \approx 53$ (MΩ/m)
$L$ = 3 meters

From the equation (1), $V_b \approx 1.1$ MV/amp, thus, in the high energy section of the Frascati Linac, the beam loading is negligible since the peak bunch electron/positron current does not exceed a few hundreds of mA.

The energy gain $U_0$, in TW/CG accelerating structures is given by [3]:

$$U_0 = \left(1-e^{-2\tau}\right)^{1/2}\left(P_{in} R_{sh} L\right)^{1/2} \quad (2)$$

Therefore, in the SLAC-type structures, the energy gained by a particle bunch travelling them through, is given by the following approximate expression:

$$U_0 \approx 10.4\sqrt{P_{in(MW)}} \qquad (3)$$

*Linac Energy Upgrading a)*

As shown in figure 4a, it is possible to upgrade the Linac energy by adding 6 new accelerating structures.

With 3 new 60 MW RF stations and by enhancing the RF pulse with the SLED energy compressor, up to 80 MW peak power can be delivered to each accelerating structure. In this case, the energy gain per section, according to the expression (3), is 93 MeV, i.e. 31 MV/m. The total extra energy is therefore 560 MeV which, added to the actual 510 MeV, gives an overall output energy of about 1070 MeV, with 5% of contingency. The space needed to install 6 new accelerating units is about 20 meters. The operation at the average accelerating field of 31 MV/m, even though feasible with the current RF technology [4], is challenging and requires much care in manufacturing, vacuum processing and RF conditioning the accelerating structures and the SLED's. Also, this version needs more powerful and costly RF stations with respect to the existing ones.

The figure 4b, instead shows the energy upgrade achieved with 12 low field (16 MV/m) new units and 2 new RF stations of the same kind of those already installed. The maximum achieved energy is 1090 MeV. This solution may be somewhat cheap and very reliable from the operational point of view. The required extra length to house 12 units is however 40 meters and the occupation of the transfer line space would be necessary.

*Linac Energy Upgrading b)*

The Linac energy doubling can be achieved, as shown in figure 5, by upgrading the accelerating field of the existing 8 units and by adding just 4 new SLAC-type sections to reach 1100 MeV. The Linac waveguide network must also be modified in order to supply two accelerating units per station. The present accelerating structures, which now operate at about 17 MV/m, will need new bake-out and RF conditioning to dissipate higher peak power. The system requires two new 45 MW klystrons. Sufficient room to install four additional accelerating structures is available in the drift space at the Linac end. The average RF field is 26.5 MV/m that is today easily attainable.

## CONCLUSIONS

Upgrading the Frascati Linac to inject at full energy in the DAFNE2 complex is feasible at moderate cost by adopting the solution b, providing new arrangement of the waveguide network and new RF conditioning of the existing accelerating sections to operate at higher field.

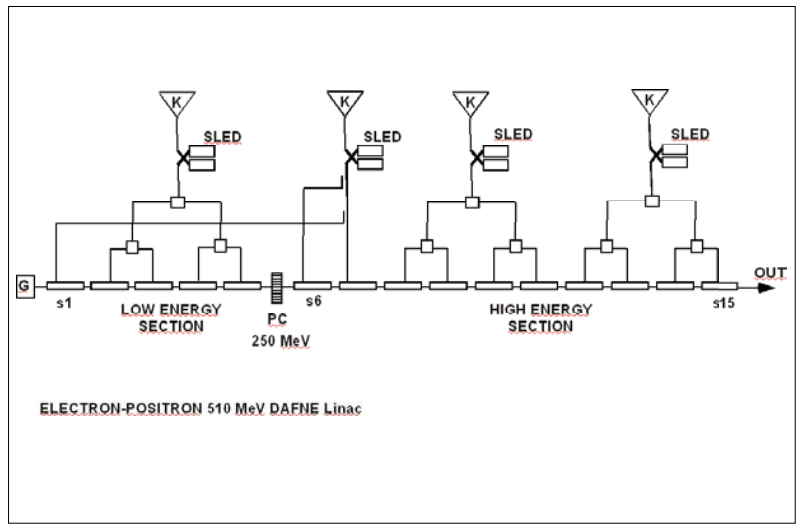

Figure 1: The DAFNE LINAC RF layout

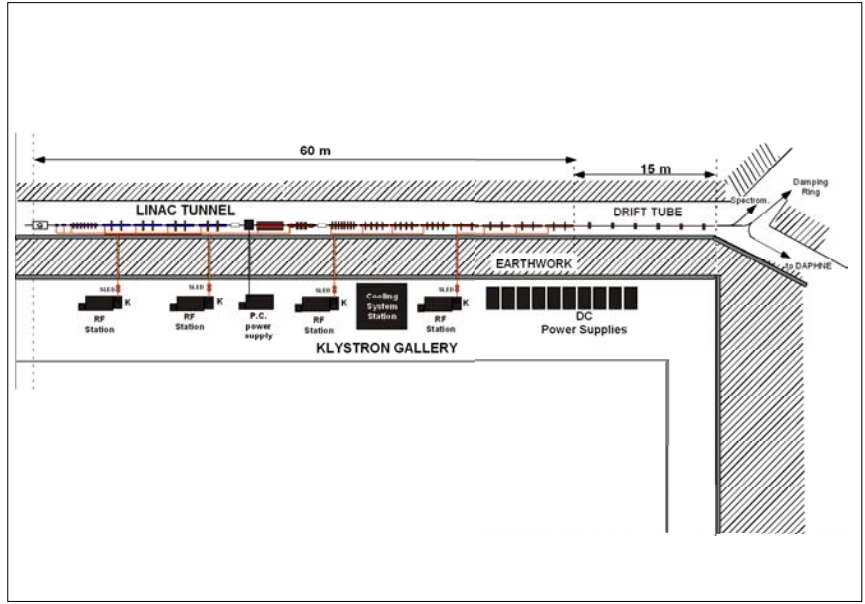

Figure 2: The DAFNE-Linac plan view

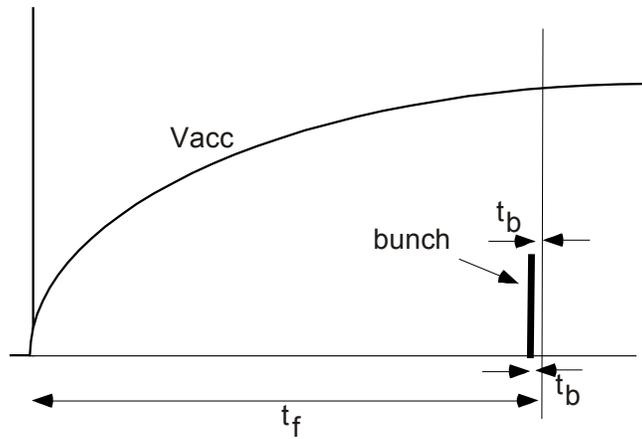

Figure 3: Accelerating structure filling voltage and bunch passage

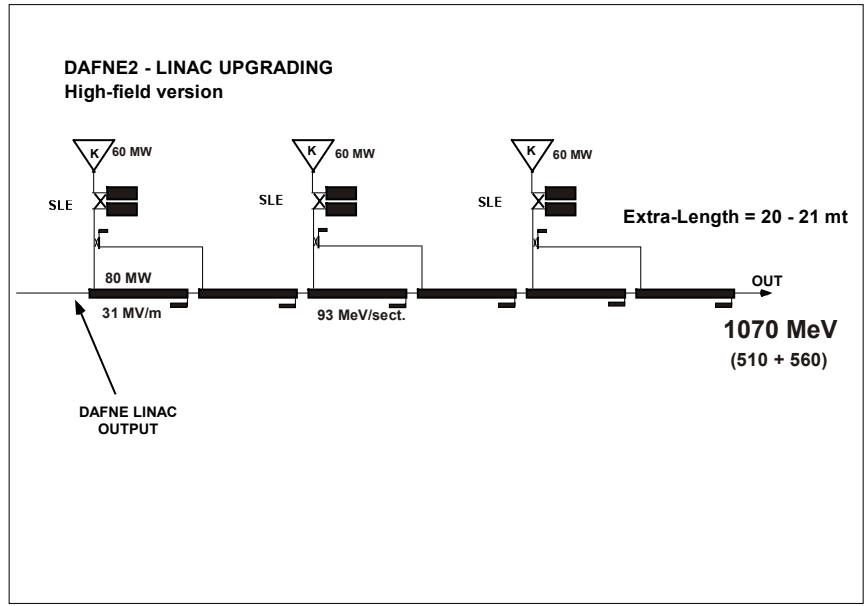

Figure 4a: Energy Upgrade with High Field Accelerating Structures

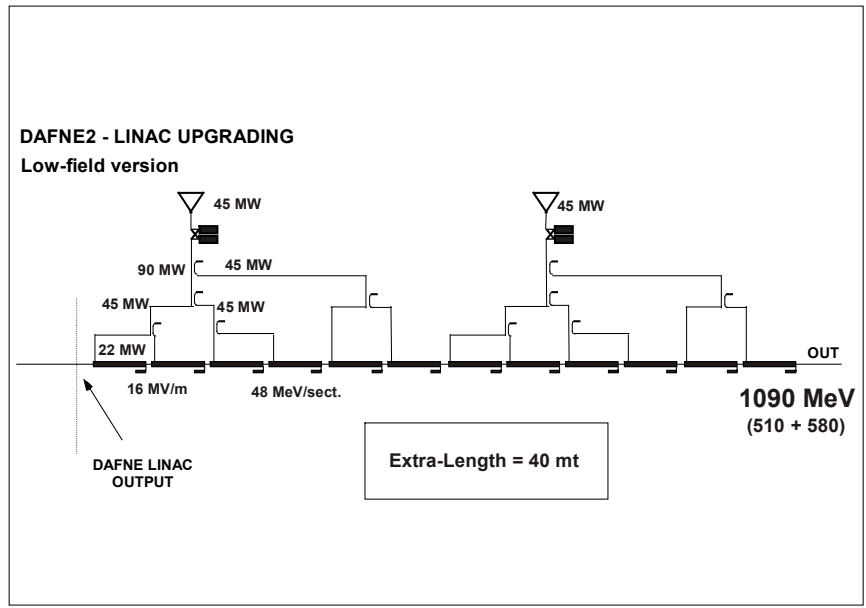

Figure 4b: Energy Upgrade with Low Field Accelerating Structures

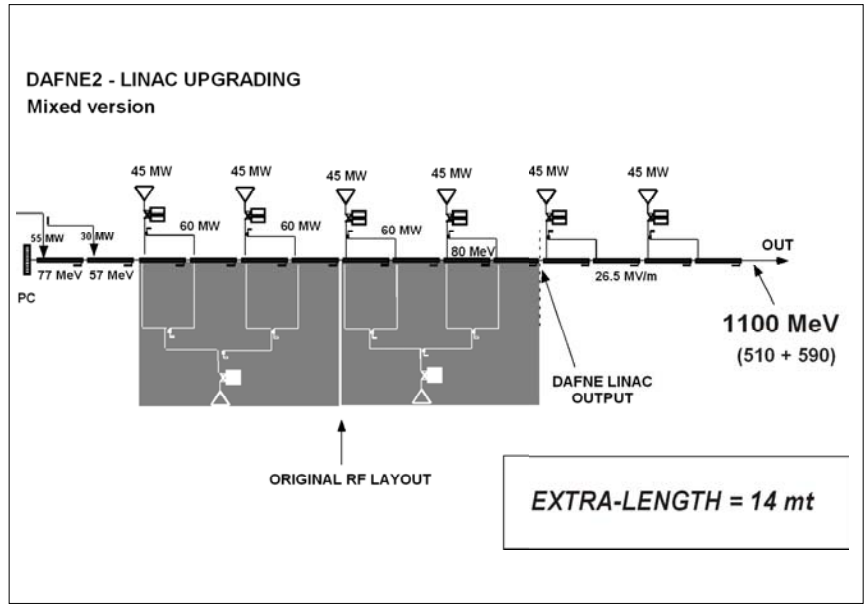

Figure 5: Linac Energy Upgrade with the new arrangement of the existing structures